\documentclass{eas}
\usepackage{graphicx}
\begin{document}

%%-----------------------------
%%      the top matter
%%-----------------------------
\title{Early evolution of tidal dwarf galaxies} 
\author{Simone Recchi$^{1, }$}
\address{Institute of Astronomy, Vienna University, 
T\"urkenschanzstrasse 17, 1180 Vienna, Austria}
\address{INAF - Osservatorio Astronomico di Trieste, 
Via G.B. Tiepolo 11, 34131 Trieste, Italy}
\author{Pavel Kroupa}\address{Argelander Institute for Astronomy, Bonn 
University, Auf dem H\"ugel 71, 53121 Bonn, Germany}
\author{Christian Theis$^1$}
\author{Gerhard Hensler$^1$}
\begin{abstract}
Our aim is to study the evolution of tidal dwarf galaxies.  The first
step is to understand whether a model galaxy without Dark Matter can
sustain the feedback of the ongoing star formation.  We present tests
of the evolution of models in which star formation efficiency,
temperature threshold, initial distribution of gas and infall are
varied.  We conclude that it is feasible to keep a fraction of gas
bound for several hundreds of Myr and that the development of galactic
winds does not necessarily stop continuous star formation.
\end{abstract}
\maketitle
%%-----------------------------
%%      your text
%%-----------------------------
\section{Introduction}

Tidal tails, resulting from the interactions of galaxies, can create
self-gravitating structures which are the seeds of dwarf galaxies.  In
these galaxies, commonly known as {\it Tidal Dwarf Galaxies} (TDGs),
the Dark Matter (DM) content is very small, which make them more
vulnerable to stellar feedback.  1-D DM-free models of dwarf galaxies
have already demonstrated that they can remain bound even with a large
fraction of supernova-expelled gas (Hensler et al. \cite{htg04}).  

\section{The model}

We perform 2-D chemodynamical simulations of DM-poor dwarf galaxies in
order to study the impact of feedback from the ongoing star formation
(SF) to such structures characterized by a low binding energy (Kroupa
\cite{k98}).  SF depends on the density and temperature of the
gas. Self-gravity and a treatment of the chemical evolution of the
stellar and gaseous component are considered.  We present here models
with an initial total gas mass of 4 $\times$ 10$^8$ M$_\odot$ and we
have modified further parameters as SF efficiency $\epsilon_{\rm SF}$,
temperature threshold T$_{\rm thr}$ for SF, initial mass distribution
(spherical vs. aspherical), and gas infall (no, subsonic, supersonic).

\section{Results}

In a model with $\epsilon_{\rm SF}=0.2$, T$_{\rm thr}$ =10$^4$ K, type
II supernovae produce a large-scale galactic wind able to unbind most
of the gas after $\sim$ 100 Myr and SF can not proceed further.  If we
reduce T$_{\rm thr}$ to 10$^3$ K, a smaller amount of gas fulfills the
conditions for the SF, therefore the galaxy does not experience a
burst but a continuous episode of SF of mild intensity lasting several
hundreds of Myr.  Also a reduction of $\epsilon_{\rm SF}$ to 0.1 leads
to a SF rate milder compared to the standard model and no disruptive
bursts occur within the first 100 Myr.  An aspherical initial gas
distribution favors the development of a large-scale outflow in the
polar direction. However, the transport of material along the R-axis
is not significant and stars still can form inside the cavity walls
(see Fig. 1).  Models with gas infall reveal that in case of a
supersonic infall a shock wave propagates towards the galaxy center
and triggers a very powerful disruptive central starburst.  A subsonic
infall provides instead a continuous reservoir of gas for the SF and
does not trigger starbursts.  Also in this case, the SF lasts several
hundreds of Myr.

\begin{figure}
  \vspace{-1.2cm}\includegraphics[height=0.7\columnwidth]
{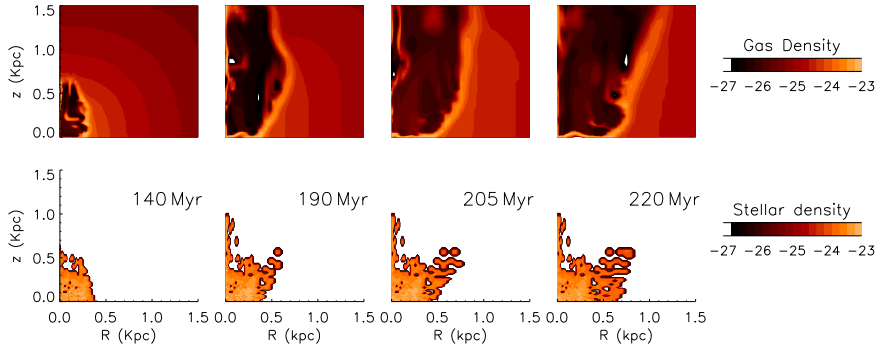} 
  \label{gs}
  \vspace{-3.6cm}
\caption{ Density
  contours of the gaseous (upper panels) and stellar (lower panels)
  phase at 4 evolutionary times for the aspherical model.}
\end{figure}

\section{Conclusions}

From our models we found that TDGs are not necessarily destroyed
rapidly and can sustain SF for times longer than 200 Myr and that
galactic winds do not always suppress SF. Further work is necessary to
sharpen our understanding of these issues.

\begin{acknowledgements}
This project is supported by the German Science Foundation (DFG) as part 
of the priority programme 1177.
\end{acknowledgements}

%%-----------------------------
%%      your bibliography
%%-----------------------------

\end{document}